\documentclass[twocolumn,showpacs,
preprintnumbers,amsmath,amssymb,floatfix]{revtex4}
\usepackage{graphicx}
\usepackage{dcolumn}
\usepackage{bm}

\newcommand{\beq}{\begin{equation}}
\newcommand{\eeq}{\end{equation}}
\newcommand{\beqa}{\begin{eqnarray}}
\newcommand{\eeqa}{\end{eqnarray}}

\newcommand{\ve}[1]{{\bm #1}}

\newcommand{\dd}{{\rm d}}

\begin{document}

\title{Classical Loschmidt echo in chaotic many-body systems}

\author{Gregor Veble$^{1,2}$ and Toma\v z Prosen$^{1}$}
\affiliation{$^{1}$Physics Department, FMF, University of Ljubljana, Ljubljana, Slovenia\\
$^2$ Center for Applied Mathematics and Theoretical Physics, University of Maribor, Maribor, Slovenia}

\date{\today}

\begin{abstract}
General theoretic approach to classical Loschmidt echoes in chaotic systems 
with many degrees of freedom is developed. For perturbations which affect essentially all degrees of freedom we find a doubly exponential decay with the rate 
determined by the largest Lyapunov exponent. The scaling of the decay rate 
on the perturbation strength depends on whether the initial phase-space density
is continuous or not.
\end{abstract}


\maketitle

Understanding macroscopic irreversibility from reversible 
miscroscopic evolution laws is one of the fundametal controversies of 
statistical mechanics, put forward by Loschmidt 
\cite{loschmidt}. The problem has been reformulated in terms of the so 
called {\em Loschmidt echo} or {\em fidelity} 
\cite{Peres,Usaj,Jalabert,Prosen,PZ},
mainly in the context of quantum chaos and quantum information, 
and only more recently \cite{PZ,Eckhardt,BenentiCasatiVeble} in the
context of classical dynamics.
In \cite{VebleProsen} it was shown that dynamics in the echo picture, which is
a composition of perturbed forward and unperturbed backward Hamiltonian 
evolution in time, can again be written as forward Hamiltonian evolution. 
The Hamiltonian in this case becomes time dependent and is
merely the perturbation part of the original Hamiltonian evaluated at a point
obtained by the forward unperturbed evolution. Analyzing the motion in the echo
picture for the case of Anosov systems 
showed that, for sufficiently long times, it can be separated into 
a composition of independent exponentially accelerated
one-dimensional motions, each such motion
locally corresponding to a Lyapunov vector direction
of the original, unperturbed system. Motion along each such (unstable) 
direction could be well described statistically, using a kernel whose shape is
time independent but whose width increases as $\sim e^{\lambda_i t}$ where 
$\lambda_i$ is the 
corresponding Lyapunov exponent. The theory \cite{VebleProsen} shows that
such an evolution leads to the Lyapunov decay of fidelity for the case of a 
system with one degree of freedom, confirming results of 
Refs. \cite{PZ,Eckhardt,BenentiCasatiVeble}. For systems with more than a 
single degree of freedom a cascade of exponential decays with increasing 
decay rates was predicted and observed.

Here we extend the results to systems with {\em many} degrees of freedom. 
While a similar theroy may still be applied, the results are significantly 
different as the regime which is considered perturbative in systems with 
few degrees
of freedom is actually the dominant one for systems with a large number of 
freedoms. What we find is that the decay of fidelity for systems with a large
number of degrees of freedom
decays as a doubly exponential function of time for the case of a {\em global}
perturbation (afftecting all degrees of freedom),
and the scaling of decay rate on the strength of perturbation depends on 
whether the initial phase space density is continuous or not. 
In another case of a {\em local} perturbation we find and justify a 
system dependent form of decay that falls between the exponential and doubly 
exponential behaviors.

In a system with $N$ phase space variables fidelity is defined as the overlap
$$
f(t)=\int \dd^N \ve x ~\rho_0(\ve x) ~\rho_t(\ve x),
$$
where $\rho_0$ is the initial phase space density, $\ve x$ is a phase space 
point and $\rho_t$ is the echo density as obtained by a composition of a 
{\em perturbed} forward evolution for time $t$ with Hamiltonian 
$H=H_0 + \delta V$, $\delta$ is the strength of perturbation,
and an {\em unperturbed} backward evolution, again for time $t$, 
with Hamiltonian $H_0$. As shown in \cite{VebleProsen}, under the 
assumption that the time is long enough in order for the echo dynamics 
to start to behave statistically, the echo density can be written as
$$
\rho_t(\ve y)=\int \dd^N \ve  y^\prime \rho_0(\ve y +\ve y^\prime) 
\prod_{i=1}^{N} \frac{1}{\sigma_i} 
K_i\left(\frac{y_i^\prime}{\sigma_i}\right),
$$
where new variables $y_i$ are introduced (see \cite{VebleProsen}) that give
the coordinates along the stable and unstable local Lyapunov vectors.
The kernels $K_i$ are time 
independent with their typical widths of the order of $1$, 
the time dependence being present in the widths 
$\sigma_i = \delta \gamma_i \exp(\lambda_i t)$ for the unstable and 
$\sigma_i = \delta \gamma_i$ for the stable directions, 
where $\delta$ is the strength
of the perturbation and $\lambda_i$ the corresponding Lyapunov exponent.
The factor $\gamma_i$ in the expression of $\sigma_i$ is determined
by the typical values of the
 phase space variable $W_i(\ve x)=\ve e_i(\ve x) \cdot \nabla V(\ve x)$,
where $\ve e_j$ is the corresponding Lyapunov vector
defined as in \cite{VebleProsen}.
Fidelity can be rewritten in the new coordinates
as $
f(t)=\int \dd^N \ve y ~\rho_0(\ve y) ~\rho_t(\ve y).
$

Let us assume now that the initial density is a product 
$\rho_0(\ve y)=\prod_{i=1}^{N} \rho_0^{(i)} (y_i)$, 
for example being a characteristic funciton on a hypercube, or a 
multidimensional Gaussian, in coordinates $y_i$.
In this case the subsequent analysis is very much simplified and we
see no reason for the results to be different in the general cases, as indeed 
the numerics, which does not correspond to initial product denisty, shows.
Under this assumption fidelity itself can be written as a product of
contributions from different individual Lyapunov directions, $f(t)=
\prod_{i=1}^{N} f_i(t)$. Each such
contribution is of the form
$
f_i(t)=\int \dd y ~\rho_0^{(i)}(y) ~\rho_t^{(i)}(y),
$
where
$
\rho_t^{(i)}(y)=\frac{1}{\sigma_i}\int \dd y'
\rho_0^{(i)}(y+y^\prime)
K_i\left(\frac{y\prime}{\sigma_i}\right)$. The initial density contributions
$\rho_0$ can be considered to be square 
normalized, and we may write 
$$
f_i(t)=1-\frac{1}{2 \sigma_i}\int \dd y \dd y^\prime 
\left[\rho_0^{(i)}(y)-\rho_0^{(i)}(y+y^\prime)\right]^2
K_i\left(\frac{y^\prime}{\sigma_i}\right).
$$
Introducing $D_i(y^\prime)=\int \dd y \left[\rho_0^{(i)}(y)-
\rho_0^{(i)}(y+y^\prime)\right]^2$ 
we may see
that the behavior of $f(t)$ will be markedly different depending on 
whether the function $\rho_0$ is continuous or not. In the continuous case
the function $D_i(y^\prime)
\propto {y^\prime}^2$ for small $y^\prime$, whereas in the discontinuous but
piecewise continuous case the behavior is 
$D_i(y^\prime)=G_i \left|y^\prime\right|$, 
where $G$ represents
the sum of all squares of the discontinuity gaps of $\rho_0^{(i)}$. 
From this it
follows that to the lowest order of $\sigma_i = \delta e^{\lambda_i t}$,
\beq
1-f_i(t)\approx
c_i \sigma_i^\beta
=
c_i \left(\delta \gamma_i \right)^\beta \exp(\beta \lambda_i t)
\label{eq:cont}
\eeq
where $\beta=1$ for the discontinuous case, 
and $\beta=2$ for the continuous case, and 
$c_i$ is a constant.

In the case of many degrees of freedom fidelity is
a product of contributions of many 
separate unstable directions. Even if individual
contributions $f_i(t)$ do not deviate significantly from $1$ that may not be 
the case for the product of all $f_i$. 
When $N$ becomes very large we may therefore 
expect the regime where equation (\ref{eq:cont}) 
is valid to describe the behavior of fidelity for all times until it becomes 
insignificantly small, at least in the cases where a significant fraction
of unstable directions contribute to the decay of fidelity.
The resulting expression is thus
\beq
f(t)=\prod_{i=1}^N \left( 1-c_i \left(\delta \gamma_i \right)^\beta 
\exp(\beta \lambda_i t)\right)
\label{eq:productfidelity}
\eeq
where the product is taken along all the unstable directions $i$, 
$c_i$ are positive
coefficients
depending on the shape of the initial density $\rho_0$ and the details of the 
perturbation $V$ and
$\beta=1,2$, depending whether the initial density is discontinuous, 
or continuous, respectively.

The behavior of fidelity depends on a few considerations. Here we
limit ourselves to the cases of initial phase space densities that are not 
constant (inpedependent) with respect to any of the phase space variables. 
Another important 
distinction is the nature of the perturbation. The perturbation can be either
local in nature, meaning that it significantly affects only a finite number
of degrees of freedom of the system, or they can be extended such that each
degree of freedom is affected equally on average. 
 
\begin{figure}[t]
\centerline{\includegraphics[width=8.5cm]{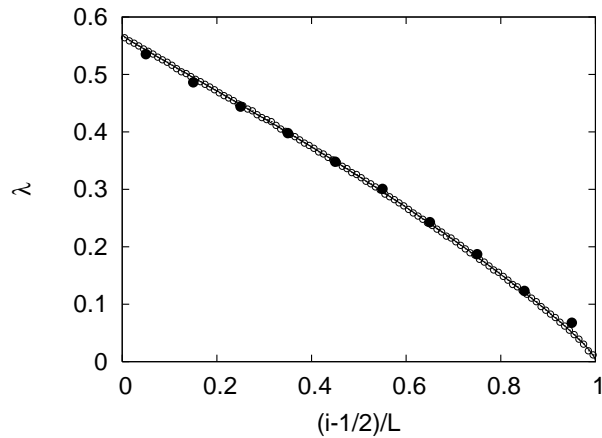}}
\caption{The Lyapunov spectrum of the kicked Ising chain 
as calculated after $10^4$ kicks
for $L=10$ (full circles), 
$100$ (empty circles) and $1000$ (line)
spins sorted from the largest to lowest positive value
(the negative and $0$ values of the spectrum are omitted). On the horizontal
axis we give $(i-1/2)/L$, where $i$ is the consecutive index of the Lyapunov
exponent.\label{fig:lyap}}
\end{figure}

In the case of an extended perturbation, each component of the
gradient of the perturbation potential $\nabla V$ is of the order of $1$, 
i.e. independent of $N$. As it may be argued that the norm of
the Lyapunov vector $\ve e_i$ is independent of $N$,
the magnitude of the 
variable $W_i$ can be estimated as
$
\left(\ve e_i \cdot \nabla V\right)^2=
\left(\sum_{j=1}^N \left[\ve e_i\right]_j
\partial_j V\right)^2\approx\sum_{j=1}^N \left[\ve e_i\right]_j^2
\left(\partial_j V\right)^2={\cal O}(1).
$
The above estimate is done by assuming that the components of the Lyapunov
field vector are not correlated. This estimate gives the magnitude of the
factor $\gamma_i$ in equation (\ref{eq:productfidelity}).
In that equation, the most relevant contributions to the decay are
those with the highest Lyapunov exponents.
As numerical evidence below suggests, for uniform systems the Lyapunov 
spectrum in the thermodynamics limit $N\to \infty$ has a limiting distribution 
of Lyapunov exponents, with an existing maximum exponent $\lambda_{\rm max}$. 
From this it follows that in an interval $[\lambda_{\rm max}-\Delta \lambda,
\lambda_{\rm max}]$ there will exist $\propto \Delta \lambda N$
 Lyapunov exponents in this interval. The number  $n$ of relevant factors
that contribute to the decay of fidelity as given in 
(\ref{eq:productfidelity})
therefore scales linearly with $N$.
Assuming that the factors $c_i \left(\delta \gamma_i\right)^\beta$ in equation 
(\ref{eq:productfidelity}) have a well defined average and are statistically
independent, when $N$ becomes very large equation 
(\ref{eq:productfidelity}) can be written as
\beq
f(t)=\exp\left(-\alpha N \delta^\beta 
\exp(\beta \lambda_{\rm max} t)\right),
\label{eq:fidext}
\eeq
where $\beta=1$ or $2$. 
The decay of fidelity in many body systems with global
perturbations is therefore 
expected to decay as a doubly exponential function, with the decay being
stronger with increasing $N$. This is the main result of this Letter.

\begin{figure}[t]
\centerline{\includegraphics[width=8.5cm]{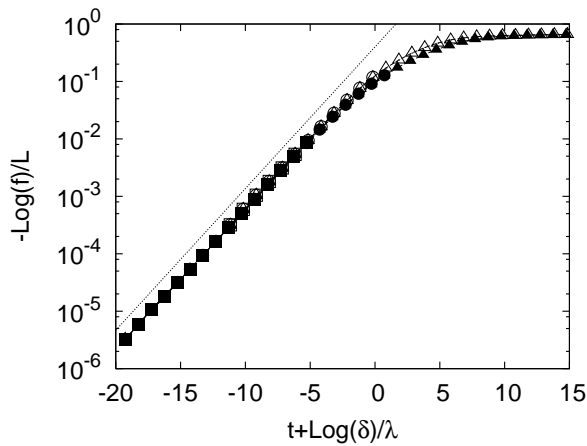}}
\caption{Decay of fidelity as a function of time for the case of the
discontinuous distribution. We divide the logarithm of
fidelity by $L$, show it in log-scale, and shift the time
by $\log(\delta)/\lambda_{\rm max}$. We show the data for $L=10$ (triangles),
$100$ (circles) and $1000$ (squares) with perturbation strengths 
$\delta=10^{-3}$ (empty symbols) and $10^{-5}$ (full symbols). The number of 
trajectories was $n_t=10^7$ for the cases of $L=10,100$ and $n_t=10^6$ for 
$L=1000$. We may see that in the regime of validity of equation
(\ref{eq:fidext}), as shown by the dotted line for $\beta=1$ with an 
arbitrarily chosen constant $\alpha$,
 all data scale to the same curve in the diagram. 
The deviations near the saturation values are due to different times of 
dynamics of the system and therefore a different decay of correlations 
(see \cite{BenentiCasatiVeble}), this regime however moves to the values
of fidelity approaching $0$ with $L\to \infty$.\label{fig:discont}}
\label{fig:2}
\end{figure}

When considering the case of a local perturbation more care needs to be taken.
Because the perturbation is local,
the variables $W_i$ that determine the spreading of densities along individual
Lyapunov directions are highly correlated amongst themselves. Namely, the 
Lyapunov vectors corresponding to the exponents close to the maximum one have
a very similar structure on the local scale (in the sense of neighbouring
degrees of freedom)
and therefore the corresponding different phase space
variables $W_i$ are, according to their definition, all very similar. 
This means that, for very
short times, the contributions to
the spreading of density along different Lyapunov directions cannot be 
considered to be independent of each other, but rather the spreading is 
effectively one-dimensional along the average of the Lyapunov directions 
in phase space. 
This average direction corresponds to the (local) 
degree of freedom in which the 
perturbation is being applied. At longer times, due to the coupling between
the neighbouring degrees of freedom, the differences between 
different Lyapunov vectors and exponents become significant
and the spreading of densities takes place in an increasing number of
dimensions.
This can be understood as the diffusion of the effect of the perturbation 
to the neighbouring degrees of freedom of the system. Thus we 
may conclude that: (i) the decay of fidelity for a local perturbation
is independent of the size $N$ of the system for sufficiently large systems, 
but (ii) that it is faster than exponential since more and more degrees of 
freedom are beginning to contribute to the decay due to the diffusion of the 
perturbation effects (exponential decay is expected if only a single degree of 
freedom would be relevant). However, since the nature of the diffusion and
the related properties of the Lyapunov vectors and the spectrum are very much
system dependent, there is no universal decay of fidelity for the case of the
local perturbation.

\begin{figure}[t]
\centerline{\includegraphics[width=8.5cm]{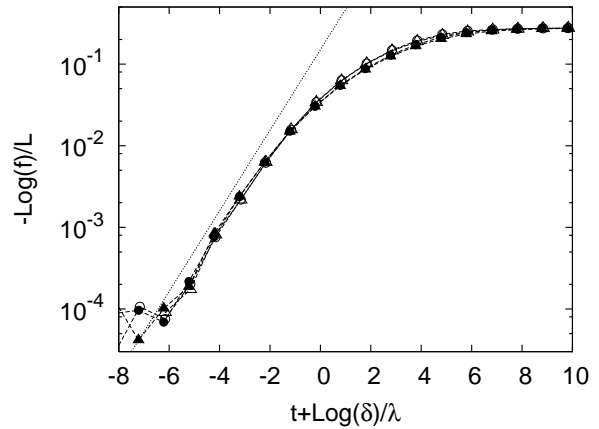}}
\caption{Decay of fidelity as a function of time for the case of the
continuous distribution, using the same represantation of scaled data as in 
Fig.~\ref{fig:2}.
We show the data for $L=10$ (triangles)
and
$30$ (circles) with perturbation strengths 
$\delta=10^{-3}$ (empty symbols) and $10^{-4}$ (full symbols). The number of 
trajectories was $n_t=
10^7$. We show the prediction (\ref{eq:fidext}) with 
arbitrarily chosen $\alpha$  and
$\beta=2$ using the dotted line.
\label{fig:cont}
}
\end{figure}

Our theoretical predictions will be demonstrated numerically in a generic
chaotic classical many-body system, namely a classical kicked Ising spin chain
with the Hamiltonian (studied in the quantum version in \cite{Prosen})
$$
H=\sum_{i=1}^{L} 
\left[J_i s_i^{\rm z} s_{i+1}^{\rm z} +\left(\sum_{j=-\infty}^{\infty} 
\delta(t-j)\right) K_i s_i^{\rm x}\right].
$$
$L$ classical spin variables $\vec{s}_i$ take values on unit spheres, 
$\vec{s}^2_i=1$, and periodic boundary conditions are assumed $L+1\equiv 1$.
The coupling and kick parameters were normally set to constant values 
$J_i=2$ and $K_i=2$, respectively. The dimension of the phase space of such
a system is $N=2L$.
The perturbation of the system was done by modifying the
kicking strength $K_i^\prime=K_i+\delta V_i$.
In the case of the global perturbation this was done for all $i$ uniformly, normally we set $V_i = 1$, 
while in the case of the local perturbation, $V_1 = 1,V_{i\neq 1}=0$. 

\begin{figure}[t]
\centerline{\includegraphics[width=8.5cm]{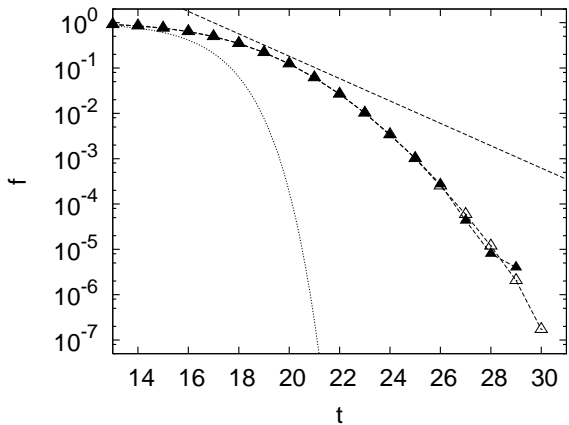}}
\caption{Decay of fidelity as a function of time when a single spin is
perturbed, and discontinuous initial density is chosen. 
We show the decay in the case of $L=100$ (empty, 
$n_t\approx 5.8\times 10^6$ trajectories) and $L=1000$
(full triangles, $n_t=10^6$) with $\delta=10^{-4}$.  
The prediction (\ref{eq:fidext})
for extended perturbations with arbitrary $\alpha$ is shown as a dotted line 
and the (arbitrarily shifted) exponential decay
with the maximum Lyapunov exponent of the system as a dashed line. 
As explained
in the text, the actual decay falls in between the two types of behavior.
\label{fig:single}}
\end{figure}

In order to demonstrate the differences between
continuous and discontinuous distributions, we used two sets of initial
densities. For the discontinuous case $\beta=1$, we used
an initial uniform distribution of all spins on the upper hemispheres 
$s_i^{\rm z}>0$. Even though the support for these distributions is not
small in length as compared to a diameter of phase space, the
fraction of the phase space volume that they cover, which gives the fidelity plateau $f(\infty)$, drops to $0$ as $2^{-L}$  in the thermodynamic limit.

For the continuous density our choice of the distribution was a product of
square normalized
distributions $w(s^{\rm z})=\sqrt{3/8}\left(s^{\rm z}+1\right)$ 
for all spins. 
This, however, does not imply that the distribution is a product distribution 
in the 
Lyapunov directions. Similarly to the case of discontinuous densities,
fidelity plateau is again expected to drop to $0$ in the large $L$ limit, 
namely $f(t\to\infty) = [2^{-1}\int_{-1}^1\dd s^{\rm z} w(s^{\rm z})]^L = 
(3/8)^{L/2}$.

To be able to test the predictions we need to determine the Lyapunov spectrum 
of such a system. We show the positive part of the Lyapunov spectrum
 in Fig.~\ref{fig:lyap}. We checked that the negative part of the spectrum
corresponds to the positive part and that the spectrum converges with time.
We may see that the 
actual density of Lyapunov exponents divided by $L$ beautifully 
scales to the limiting density when $L$ is large enough. 
Such a behaviour is 
typically found for interacting many-particle systems, such as e.g. 
Fermi-Pasta-Ulam chain, Lorentz gas, etc.
This justifies our assumption that the number
of Lyapunov exponents in the neighbourhood of the largest Lyapunov exponent
 increases linearly with $N=2L$, 
yet the maximum Lyapunov exponent converges with $N$. 

We show the decay of fidelity for the discontinuous case and a global 
perturbation in Fig.~ \ref{fig:discont}. Using the appropriate scaling with respect to
the number of spins $L=N/2$ and perturbation strength $\delta$ as predicted
in equation (\ref{eq:fidext}), where $\beta=1$, we may see that the 
doubly exponential behavior with the predicted dependence on $\delta$ and
$N$ holds true. It should be noted that, while there is an observed plateau
of fidelity when $N$ is still small, this plateau exponentially tends to $0$
with $N$ and indeed the doubly exponential behavior of fidelity can be 
observed for all values of fidelity for large $N$. 

Similar observations hold true for the continuous case where $\beta=2$ as
shown in Fig.~\ref{fig:cont}. For
the same system, using continuous densities results in a faster form of
decay with time and with quadratic dependence on the perturbation.
In the continuous case we also observe expected numerical
statistical fluctuations 
of fidelity at values close to $1$, especially for large $N$. This is because
the initial density is represented by a finite number of trajectories and even
when $t=0$ the estimation of fidelity will differ from $1$. Nevertheless,
the predicted scaling with $N$ and $\delta$ is observed along
with the faster decay of fidelity with time as compared to discontinuous case.

For the case of a local perturbation with discontinuous densities as
presented in Fig.~\ref{fig:single}, 
we observe a decay that lies in between
the exponential behavior, as expected in systems with a single degree of
freedom, and the doubly exponential behavior as is found in the global
perturbation case.
Details of such behavior are
system dependent as they are determined by the diffusion of the effects of
the perturbation to neighbouring degrees of freedom.

In addition we have tested our predictions for the disordered cases
where the kicking strengths $K_i$ randomly varied from spin to spin, as well as
the perturbations of the kicking strength $V_i$, however the results were
again 
confirming our theory.

In conclusion, we have shown and explained that the decay of the classical 
Loschmidt echo in systems with a large number of degrees of freedom decays
as a doubly exponential function for a global perturbation, 
and the decay rate depends on whether
the phase space densities chosen are continuous or not. For the case of a 
local perturbation a system dependent form of decay is observed and justified
that falls between exponential and doubly exponential behaviors.
We acknowledge support by the grant P1-044 of Ministry of Science and 
Technology of Slovenia, and in part by the U.S. ARO grant DAAD19-02-1-0086.

\end{document}